# Expansion microscopy reveals neural circuit organization in genetic animal models

Short title: **Expansion microscopy for neural circuits**

**Authors:**
Shakila Behzadi [1], Jacquelin Ho [2], Zainab Tanvir [3], Gal Haspel [1*], Limor Freifeld [4*], Kristen E. Severi [1*]

1. Federated Department of Biological Sciences, New Jersey Institute of Technology, NJ, USA
2. Departments of Genetics and Neuroscience, Albert Einstein College of Medicine, NY, USA
3. Departments of Cell Biology and Neuroscience and of Genetics, Rutgers University – New Brunswick, NJ, USA
4. Faculty of Biomedical Engineering, Technion - Israel Institute of Technology, Israel

*co-corresponding authors
*GH: haspel@njit.edu, LF: freifeld@bm.technion.ac.il, KES: severi@njit.edu*

## ABSTRACT

Expansion Microscopy is a super-resolution technique in which physically enlarging samples in an isotropic manner increases inter-molecular distances such that nano-scale structures can be resolved using light microscopy. This is particularly useful in neuroscience as many important structures are smaller than the diffraction limit. Since its invention in 2015, a variety of Expansion Microscopy protocols have been generated and applied to advance knowledge in many prominent organisms in neuroscience including zebrafish, mice, *Drosophila*, and *C. elegans*. Here we review the last decade of Expansion Microscopy-enabled advances with a focus on neuroscience.

## INTRODUCTION

Until recently, important neuronal structures have been too small to resolve with light microscopy. Resolution in microscopy is the capacity to distinguish closely spaced objects as separate entities (Dekker & Bos, 1997). The resolution of light microscopy is physically limited by the wavelength of light and the collecting angle of the front lens to approximately 200 nanometers, known as the 'diffraction limit' (Abbe, 1873 ; Betzig & Trautman, 1992). Many neuroanatomical structures, such as chemical synapses, gap junctions, and the smallest of neurites, are smaller than the diffraction limit. The two solutions for this issue have been to use either electron microscopy, in which the diffraction limit is dictated by a much smaller wavelength that allows molecular level resolution, or innovative modifications of optics and image analysis in a suite of approaches known as super-resolution light microscopy (Arizono et al., 2023; Hugelier et al., 2023; Werner et al., 2021). Expansion microscopy (ExM) is a recent super-resolution microscopy technique that offers practical and scientific advantages (Chen et al., 2015). The four primary advantages of ExM are: 1) nano-scale resolution using standard, accessible, and broadly available diffraction-limited microscopes; 2) rapid imaging and processing time in comparison to electron microscopy and most other super-resolution methods; 3) relatively low cost; and 4) an ability to image large tissue samples at resolutions beyond the diffraction limit.

ExM increases the resolution of standard fluorescent microscopes beyond the diffraction limit by physically expanding the sample. The samples are embedded in polyelectrolyte hydrogels that isotropically expand upon the addition of pure water. Once expanded, features smaller than the diffraction limit may become resolvable using standard diffraction-limited microscopes.
Thus, the method is compatible with existing laboratory workflows.

By changing the sample rather than the optics or illumination protocol, the ExM approach differs from that of other super-resolution methods, including stimulated emission depletion (STED) microscopy, structured illumination microscopy (SIM), photoactivated localization microscopy (PALM), and stochastic optical reconstruction microscopy (STORM) (Huszka & Gijs, 2019). A limitation these techniques have in common is that they can only be used with thin sections (Vicidomini et al., 2018). STED requires a precisely aligned dual-laser setup: one laser excites a focal area, while a second laser selectively depletes fluorescence in a toroid region around the focal point to achieve super-resolution (Hell & Wichmann, 1994). SIM enhances resolution by projecting patterned light onto the sample for computational reconstruction. This method is limited to doubling the optical resolution, and it requires sophisticated light systems and precise control over illumination patterns (Gustafsson, 2000). PALM and STORM both rely on the stochastic activation and precise localization of individual fluorescent molecules to build high-resolution images over time (Betzig et al., 2006; Rust et al., 2006). PALM uses photoactivatable fluorescent proteins, while STORM typically uses photoswitchable dyes. Both methods require sensitive detection equipment, such as high-performance cameras, and advanced computational algorithms for image reconstruction (Almada et al., 2015). In addition, they require long acquisition times as each imaged plane must be imaged multiple times to allow the reconstruction of the structural information from single-particle localization data. While all these techniques achieve resolution below the diffraction limit, they require significant investment in specialized hardware and software and limit the sample size that can be imaged.

ExM, by contrast, offers a more accessible and scalable approach, leveraging broadly available microscopy equipment to study large biological specimens with high resolution. The enhancement of resolution is uniform along all three imaging planes, and the sample processing results in an entirely clear sample, which obviates the need for additional clearing techniques. The sample preparation for ExM can often be simply incorporated into existing workflows with the addition of a few days of processing time in exchange for dramatically improved imaging results. However, ExM requires fixed samples, while other methods mentioned above can be compatible with live super-resolution microscopy.

ExM permits mechanistic insights at multiple levels of organization, from ultrastructure to tissue, such as the organization of proteins within and between gap junctions and chemical synapses. Here, we review how ExM has been applied to reveal previously unknown features of neural circuits in model animals used in neuroscience research.

**<u>Expansion Microscopy</u>**

The central feature of ExM is the isotropic expansion of tissues that have been embedded in a polyelectrolyte gel. The gel expands equally in each dimension, retaining spatial relationships between features in the original tissue. Importantly, it is possible to choose which molecules will be anchored to the gel (e.g., proteins, nucleic acids, or lipids), and thus, different types of cellular structures can be captured. The standard protocol attaches the amine moieties of proteins to the gel (see proExM below). ExM typically expands a sample by about fourfold (Chen et al., 2015; Tillberg et al., 2016), but appropriate optimization of the gel composition can yield tenfold single-step expansion (Damstra et al., 2022; Truckenbrodt et al., 2018), and iterative expansion approaches can be used to achieve expansions of about twentyfold (Chang et al., 2017; Sarkar et al., 2022). A recent protocol achieves 20x single-step expansion through further optimizing gel composition and the polymerization environment (Wang et al., 2024). Samples imaged with ExM can be stained with fluorescent antibodies both before or after the expansion step (Asano et al., 2018).

Isotropic expansion via hydration of a sample embedded in a hydrogel involves four major steps (Figure 1, Perelsman et al., 2022). The first step is anchoring, in which the sample is incubated in a solution containing an anchoring agent (Figure 1.1). The anchoring step attaches the sample to the polymerizing hydrogel. A key innovation is the use of acryloyl-X (AcX) as an anchoring agent. Incubation of the sample in AcX enables the crosslinking of amines in proteins and other molecules to acrylamide monomers in the gel. As a result, native proteins are retained - this version of ExM is known as protein-retention ExM (proExM, Tillberg et al., 2016). The second step is gelation, in which the sample is bathed in monomers (sodium acrylate and acrylamide) and a cross-linking agent that first diffuse into the sample at a low temperature and are then allowed to polymerize to form the gel (Figure 1.2). The third step is digestion, in which the sample is chemically homogenized via the application of proteolytics and detergents that disrupt the chemical bonds of molecules in the sample (Figure 1.3). Homogenization is necessary to avoid distortion of the sample during the expansion phase, while the anchoring preserves the relative locations of fluorophores that mark the structures of interest within the sample. Finally, the sample is expanded by placing it in water (Figure 1.4).

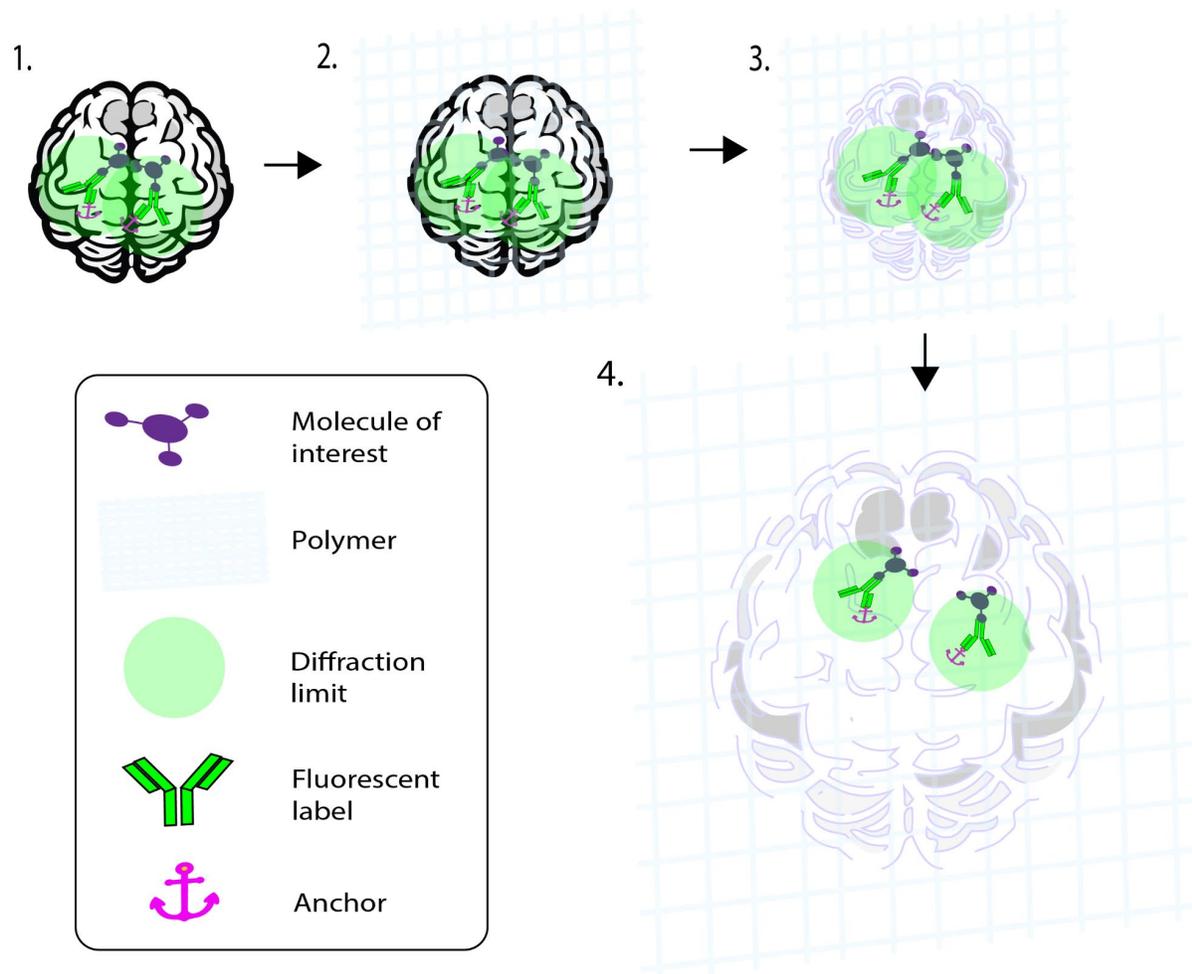

**Figure 1. Major steps of an expansion microscopy protocol.** (1) Anchoring biomolecules within the sample to gel elements. (2) Gelation of the sample and forming the hydrogel polymer. (3) Digestion of the sample through proteolysis. (4) Expansion of the hydrogel with water, enhancing resolution by spatial separation of the fluorescent labels, surpassing the diffraction limit of ~250 nanometers for light microscopy.

**TISSUE PREPARATION**

**Anchoring links sample components to the hydrogel monomers.**

Currently, most ExM relies on AcX as an anchor (Tillberg et al., 2016), which enables crosslinking of amine moieties in proteins and other molecules to acrylamide monomers in the gel. Anchoring with AcX is compatible with conventional fluorescent labeling techniques, including immunohistochemistry. Importantly, this method is compatible with both genetically expressed fluorescent proteins and commercially available antibodies, and nucleic acids when using an additional anchoring agent, such as Label-IT, to covalently attach amines to nucleic acids (Chen et al., 2016). We note, however, that some fluorescence loss can occur in the digestion and polymerization steps and the extent to which signal is lost depends on the fluorophore chemistry.

In particular, cyanine-based dyes (such as Alexa 647, cy2, cy3, and cy5) are associated with significant signal loss in proExM and thus are not recommended (Tillberg et al., 2016). N-acryloxysuccinimide (AX) has recently proven to be an anchoring agent similar to AcX and is available at much reduced cost (Wang et al., 2024).

**Gelation polymerizes and solidifies the hydrogel in which the sample is embedded.**

Gelation results in a 3D matrix of crosslinked polymers to which molecules of interest in the sample are chemically anchored. The gelation process occurs in a chamber that holds the sample and the solutions that form the gel. The chamber must be large enough to cover the sample in its entirety and establish the shape of the gel. The chamber is then sealed with a coverslip and incubated to allow polymerization (Perelsman et al., 2022). The size and shape of the chamber are additional critical features that determine the user's ability to manipulate the gel and its structural integrity during processing. After the monomer solution diffuses throughout the sample in a 4°C environment, a change to 37°C triggers the polymerization of the sodium acrylate and acrylamide monomers (Gao et al., 2021). The linear polymers then form crosslinks to create a dense matrix and through the anchoring molecule covalent bonds form between amines in the sample and acrylamide in the gel.

Most existing protocols for ExM depend on bis-acrylamide, which cross-links acrylamide monomers in the gelation step. N,N-dimethylacrylamide (DMAA) is a self-polymerizing agent which forms more structurally robust hydrogels that are able to withstand 10x or even 20x expansion (Klimas et al., 2023; Wang et al., 2024). However, DMAA is even more sensitive to atmospheric oxygen, which reacts with intermediate radicals and interferes with radical-dependent gel cross-linking. Nitrogen gas can be used to flush out ambient oxygen from a working environment for uniform gelation (Wang et al., 2024).

**Digestion homogenizes the sample through proteolysis to minimize the distortion of the sample during later expansion**.

Digestion is typically performed using proteinase-K (Tillberg et al., 2016). For experiments in which labeling is performed after the expansion step, this step can be replaced with autoclaving the samples in detergents such as SDS and Triton, or via the use of a milder protease, to preserve epitopes while denaturing protein aggregates to allow gel expansion (Asano et al., 2018).

**The sample-embedded hydrogel absorbs water and expands.**

Water is attracted to the polarized ends of the hydrogel molecules and forms dipoles around the negative ionic charges in the gel mesh, causing the polymer chain molecules to extend and the gel to swell isometrically (Hümpfer et al., 2024). Samples typically expand approximately fourfold (although see additional variants described below).

**<u>Expansion Variations</u>**

**High Expansion factors and resolution**

One approach to obtain expansion factors higher than approximately fourfold, and thus higher spatial resolution, is the iterative application of the expansion process. Iterative Expansion (iExM) involves two consecutive expansion rounds. In iExM (Figure 2), an expansion factor of ~20 fold is obtained, yielding a resolution of 20-25 nm with standard confocal microscopy imaging (Truckenbrodt et al., 2018).

In an alternative approach, 10X Expansion and TREx (Figure 2), to increase the expansion factor, 10-fold expansion can be achieved in a single step via optimization of the gel recipe (Damstra et al., 2022; Truckenbrodt et al., 2018).The most recent improvement of this permits 20x in a single-shot expansion (Wang et al., 2024), by developing a super-absorbent hydrogel recipe, tightly controlling oxygen levels during gelation and optimizing the gelation times according to the biological sample type (Figure 2).

**Compatibility with light microscopes: tradeoffs and considerations**

When working with large, expanded samples, the working distance between the imaged plane and the microscope objective becomes an important consideration. Working distance depends on the sample, expansion factor, and the microscope. Water-dipping objectives and upright microscopes offer the best working distances and are thus advantageous over inverted microscopes and oil or water immersion objectives. However, inverted microscopes are in more common use and the latter objectives are associated with higher numerical apertures. Superficial target molecules within the sample may be accessible with objectives that have working distances smaller than the gel thickness. Consequently, the orientation of the sample within the gel and the orientation of the gel are important when imaging. It is important to fit the chamber height as precisely as possible to the sample depth to minimize empty space (Perelsman et al., 2022).

**Labeling target proteins and nucleic acids**

To pinpoint spatial relationships among molecular structures, proteins and nucleic acids are tagged with fluorescent molecules or other markers that have high specificity to the molecules under study. ExM is typically paired with protein and nucleic acid labeling techniques, such as immunohistochemistry and *in situ* hybridization, to tag molecules of interest. By anchoring the molecules of interest to the gel using AcX for proteins, or Label-IT together with AcX for nucleic acids, and then expanding the sample, ExM enhances the detection of labeled proteins and nucleic acids, making it easier to study complex and densely packed structures within cells and tissues (Campbell et al., 2020; Chen et al., 2016; Tillberg et al., 2016; Wang et al., 2024).

**Pre vs post-expansion labeling**

Molecules of interest can be labeled either before or after expansion treatment depending on the scientific question and preferred imaging tool. Pre-expansion labeling is most common in established protocols and procedures; fluorophores are often better able to withstand the digestion step necessary to soften and homogenize samples for expansion, whereas native proteins may be proteolyzed and unfit for labeling post-expansion. However, post-expansion labeling is ideal for densely packed targets that benefit from "decrowding", an increase in target accessibility that has helped visualize nanostructures such as amyloid plaques of Alzheimer's disease model mice samples (Sarkar et al., 2022). In specific cases, since the expansion process physically separates proteins, it can render previously inaccessible targets accessible for antibody labeling, as native inter-protein distances can be shorter than the size of antibodies (Sarkar et al., 2022; Wang et al., 2024). Moreover, it can allow for more precise localization of molecules of interest by reducing the linkage error, an error in spatial visualization measured by the distance between the molecule of interest and the fluorescent reporter due to the size of antibodies used as reporters (Hamel & Guichard, 2021); the relative size of the antibody with respect to the structure of interest becomes significantly smaller when the antibody is applied to the expanded sample. Additionally, post-expansion labeling avoids fluorescence signal loss during homogenization and polymerization, which can be significant (Shi et al., 2021). Although post-expansion labeling allows for flexibility in experimental design and has greater signal amplification, it requires a gentler digestion step to retain target isotopes and greater reagent quantities and thus costs more than pre-expansion labeling due to the increased size of the sample. To mitigate the signal loss during the expansion step, pre- and post-expansion labeling can further amplify the labeling signal (Yu et al., 2020; Wang et al., 2024).

**Limitations of ExM**
Several technical challenges remain in ExM, which future advancements aim to address. One such challenge is consistency in protocol application, and obtained expansion factors, between samples of the same type and to a greater extent in different types of samples. Moreover, ExM protocols, particularly ones associated with high expansion factors, can involve complex steps, and require adaptation for and validation in new types of samples. Nevertheless, recent developments in automated sample preparation and streamlined protocols have shown promise in reducing benchwork time and improving workflow efficiency.

Additional challenges arise from the size of the expanded samples, particularly when expansion factors are 10x or higher. This is a challenge for the imaging itself, as some microscope imaging chambers have limited sizes, high NA objectives have limited working distances and thus are ill suited for imaging these samples – giving rise to a resolution limit (albeit this limit is far more then compensated by the expansion factor), and large samples require the use of large working distance objectives that poorly fit, e.g., inverted microscopes. Moreover, sample volumes can increase by 3 orders of magnitude, which can make imaging times very long even on fast microscopes. Light-sheet and upright spinning disk confocal microscopes are thus best suited for imaging expanded samples. Large sample volumes also dilute the fluorescent signal, requiring signal amplification or the use of post-expansion staining which requires very large amounts of antibodies (see pre- vs post-expansion labeling). Finally, as ExM achieves higher resolutions and sample volumes increase, the size of resulting datasets grow substantially, increasing storage and computational demands and requiring the development of appropriate software solutions for data analysis (Gao et al., 2019).

**EXPANSION MICROSCOPY ACROSS SPECIES**

Expansion microscopy has been applied to many commonly used model systems, including zebrafish (*Danio rerio*), mice (*Mus musculus.*), fruit flies (*Drosophila melanogaster*), and nematode worms (*Caenorhabditis elegans*). ExM enhances microscopic details while preserving the overall spatial structure, which is particularly useful for understanding neural structures. In zebrafish, proExM has detailed synaptic connections (Tillberg et al., 2016). In mice, ExFISH has mapped deep brain regions (Chen et al., 2016). For *Drosophila*, ExLLSM has improved imaging speed and clarity (Gao et al., 2019). In *C. elegans*, ExM has revealed protein arrangements and centriole dynamics (Suen et al., 2023). In each of these systems, several ExM variations have been used to answer fundamental questions about neuroanatomical organization at the subcellular level.

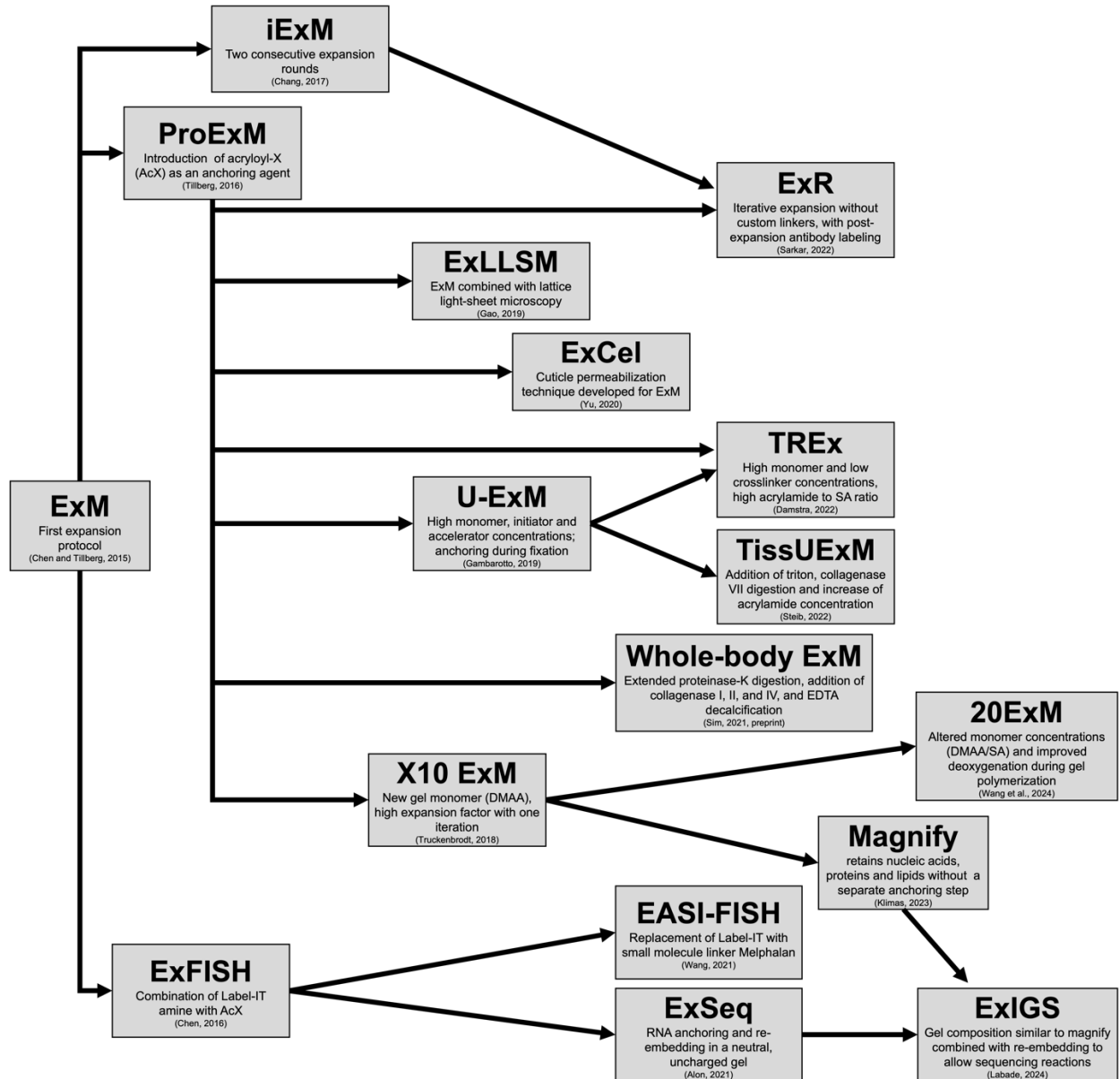

**Figure 2. Variations of Expansion Microscopy. ExM:** In the original protocol, the monomer gel solution includes eight components - sodium acrylate, acrylamide, NN'-methylenebisacrylamide. MBAA; Phosphate-buffered saline, PBS; ammonium persulfate, APS; N,N,N',N'-Tetramethylethylenediamine, TEMED; and 4-hydroxy-tempo, 4HT; Sodium acrylate and acrylamide form the polymer as they are cross-linked by MBAA. 4HT is a polymerization inhibitor that prevents premature gelation before the monomer solution diffuses throughout the sample tissue; TEMED serves as an accelerator; APS serves as the initiator and thus is added last to the solution (Chen et al., 2015; Klimas et al., 2019). **ProExM:** Acryloyl-X is added as an anchoring

agent that binds proteins to the polymer (Tillberg et al., 2016). **iExM:** An iterative approach to ExM, where the initial polymer is broken down, and a second swellable polymer mesh is formed in the space newly opened by the first expansion. This was implemented using custom-made DNA oligonucleotides conjugated to secondary antibodies and polymer anchoring moieties (Chang et al., 2017), such that only the specifically labeled structures of interest are attached to the plymer. ExR: AcX was applied in iExM for broad anchoring of proteins and antibodies to the polymer before post-expansion staining (Sarkar et al., 2022). **10X Expansion:** A higher expansion factor was obtained by replacing acrylamide with N,N-dimethylacrylamide acid (DMAA) as the main monomer together with sodium acrylate (Truckenbrodt et al., 2018). **20ExM:** A single-step twentyfold expansion protocol without iterative steps, using optimized hydrogel integrity and oxygen-controlled gelation (Wang et al., 2024). **TREx:** By decreasing the crosslinker and monomer concentrations, the hydrogel maintains structural integrity and increases the expansion factor tenfold in a single gelation step. (Damstra et al., 2022). **U-ExM**: For expanding cultured, cells and analyzing proteins at higher resolution. Anchors proteins to the polymer by applying acrylamide and formaldehyde rather than using AcX (Gambarotto et al., 2019). **iU-ExM:** Enables high-resolution visualization of cellular structures by combining ultrastructure expansion with iterative expansion, achieving near-SMLM resolution for detailed molecular imaging in tissues and organelles (Louvel et al., 2023). **TissUExM:** To penetrate thicker tissue, this protocol adds triton to the monomer solution and increases the concentration of acrylamide. Additionally, a collagenase VII digestion step is introduced after gelation, followed by a denaturation step and post-expansion staining (Steib et al., 2023; Steib et al., 2022). **Whole-body ExM:** to homogenize and expand an entire zebrafish with multiple tissue types such as bones, cartilage, muscles, and soft tissues, this protocol extended the proteinase-K digestion time, added collagenase I, II, IV digestion, and decalcification with ethylenediaminetetraacetic acid (EDTA) (Sim et al., 2021, preprint). **ExFISH:** Preserves the integrity and spatial positions of RNA molecules in cells and tissue slices. The RNA is stained by fluorescence in-situ hybridization (RNA-FISH) amplified with hybridization chain reactions (HCRs) that enable resolving single RNA molecules. Multiple rounds of FISH increase the number of target RNA sequences (Chen et al., 2016). **EASI-FISH:** Replaces Label-IT in the ExFISH protocol with a smaller molecule linker, melphalan, to retain and anchor nucleic acids to the ExM polymer. This linker is cost-effective, improves signal-to-noise ratio, and optimizes reagent penetration (Wang et al., 2021). **ExLLSM:** Addresses microscopy working distance that limits sample thickness with a light sheet microscopy pipeline (Gao et al., 2019; Lillvis et al., 2022). ExCel: Adapted ExM to the nematode C. elegans by adding steps for cuticle permeabilization (Yu et al., 2020; Yu et al., 2022). **Magnify:** Achieves higher expansion factors and enhanced imaging resolution by adding N,N-dimethylacrylamide (DMAA) to the monomer solution alongside acrylamide and sodium acrylate. This protocol combines the high expansion factor of 10× Expansion with protein retention strategies from ProExM, allowing for detailed imaging of proteins in large or thick samples (Klimas et al., 2023). **ExSeq:** Enables spatially resolved RNA sequencing by anchoring RNA molecules to the gel using reagents like LabelX and maintaining a neutral pH during expansion to preserve RNA integrity. The protocol involves re-embedding the expanded gel in a non-expandable gel to facilitate sequencing procedures (Alon et al., 2021). **ExIGS:** Facilitates in-gel sequencing of DNA or RNA molecules with higher expansion factors by using a modified polyacrylate gel incorporating DMAA and Tn5 transposase. By adding rolling circle amplification (RCA) and sequencing-by-synthesis (SBS) buffers with increased gel porosity, it supports enzymatic diffusion necessary for sequencing. The protocol includes re-embedding the expanded gel in a non-expandable gel for sequencing procedures and is derived from Magnify and ExSeq (Labade et al., 2024).

**Macroscopic to ultrastructural resolution in zebrafish**

Zebrafish (*Danio rerio*) are widely used in biomedical research partly due to their genetic accessibility, small size, and translucent embryos and larvae. Embryos and larvae are typically less than 4 mm in length and are sufficiently transparent that their internal organs can be visualized in intact, living individuals (Antinucci & Hindges, 2016). This transparency makes zebrafish particularly amenable to *in vivo* optical techniques, including neural activity imaging via genetically-encoded calcium indicators using a variety of microscopy approaches (Ahrens et al., 2012; Higashijima et al., 2003; Kim et al., 2017; Lauterbach et al., 2015; Migault et al., 2018; Renninger & Orger, 2013), optogenetic perturbations of neuronal activity, and other optical processes (Piatkevich & Boyden, 2024; Portugues et al., 2014; Simmich et al., 2012; Varady & Distel, 2020). Many neural circuits span large areas of the body, and neurons and their processes can be easily visualized throughout the nervous system in young zebrafish. Thus, zebrafish are well-suited for the study of how intact neural circuits give rise to functional behavior. However, the synaptic interfaces among neurons, critical for understanding neuronal circuit function and behavior, are densely packed with proteins that are difficult to visualize using light microscopy. Visualizing and resolving synaptic protein organization below the diffraction limit can increase our understanding of mechanisms of neural circuit function, structure, development, and plasticity.

To understand circuit function, it is advantageous to study simultaneously the macroscopic structures and nanoscopic details. ExM makes this possible by maintaining the overall spatial relationships of cellular-level structures while improving the resolvability of nanoscopic structures. ExM has been used in zebrafish to resolve synaptic connectivity and intra-synaptic protein organization, such as glycine receptors and gap junction organization within densely packed synapses using protein-retention ExM (proExM) (Cárdenas-García et al., 2024; Freifeld et al., 2017; Tillberg et al., 2016). ExM has also been used in zebrafish to resolve structures in large tissue samples such as whole-embryo zebrafish in TissuExM and whole-body expansion (Sim et al., 2021, preprint; Steib et al., 2023; Steib et al., 2022).

**Resolving synaptic connections in zebrafish**

The first demonstration of the utility of ExM in zebrafish came in Freifeld et al., 2017, where the authors resolved putative synaptic connections between two fluorescently labeled cell populations. This structural information could be highly valuable in complementing functional data from such neuronal populations thought to constitute a neural circuit. E.g., it can reveal how the participating neurons and connectivity patterns vary between individuals to mediate heterogeneity in behavior or how they change in learning and development (Freifeld et al., 2017). One such circuit in zebrafish which has been a focus of research for decades is the escape circuit. Like most fish and amphibian species, zebrafish have a pair of large command neurons for escaping predation, the Mauthner cells (Hale et al., 2016). The subcellular spatial distribution of electrical and chemical synaptic inputs onto Mauthner cells, which ultimately determine the initiation of escape behavior, are below the diffraction limit (Eaton et al., 2001; Yao et al., 2014; Zottoli et al., 1987). Several studies using ExM have resolved organizational aspects of synaptic structure and revealed insights into the mechanisms of Mauthner cell function.

Glycinergic inhibition onto the Mauthner cells is an essential gating mechanism controlling their activity and thus modulating escape response thresholds and directionality (Koyama et al., 2011). Researchers have used proExM to image larval zebrafish brains and resolve protein organization within synapses on the Mauthner cells. In particular, glycine receptors in glycinergic synapses on the Mauthner cells were found to often form annuli and the density of synaptic proteins in axon-cap synapses was found to be heterogeneous (Freifeld et al., 2017). The annular organization of glycine receptors on larval zebrafish Mauthner cells was consistent with the identified structure of glycinergic synapses in other examples, including mammals (Alvarez et al., 1997; Seitanidou et al., 1988; Triller et al., 1990). This work paves the way to reveal how this organization is functionally modulated during synaptic plasticity. ProExM can be used to answer questions that functionally link nano-scale synaptic plasticity to circuit function and behavior, since the structure can be captured subsequently to functional experiments in this animal model.

In another study researchers mapped the distribution of proteins at a single synaptic input to the Mauthner cell to reveal the structural support provided by the molecular scaffold underlying gap junctions (Cárdenas-García et al., 2024). Cárdenas-García *et al*. found multiple gap junctions and components of adherens junctions occupying most of the synaptic area, suggesting a functional association between these structures. These synapses are known to be mixed electro-chemical synapses. However, glutamate receptors were confined to peripheral portions, indicating that the majority of the synaptic area functions as an electrical synapse (Cárdenas-García et al., 2024). By identifying gap junctions at a single synapse level, electric synapses on the M-cell can be used as a model to study plasticity and protein-specific synapse configurations with functional implications. Moreover, ExM can shed light on how chemical and electrical synapses are co-modulated by experience to mediate learning.

ExM was also used to resolve the spatial relations between neural fibers and glial processes (Mu et al., 2019). This built upon a previous demonstration of ExM-enhanced tracing of radial glial cells in visual processing areas (Freifeld et al., 2017). Since glia are important for synaptic function and plasticity, resolving such spatial relations can lead to key insights into the underlying mechanisms of neural-glia interactions and glial effects on neural circuit functions and behavior.

**Resolving structures within whole intact zebrafish with ExM variants**

In Tissue Ultrastructure Expansion Microscopy (TissUExM) and Whole-body ExM (Figure 2), protocol advancements, and in particular added steps for digestion of collagen and bones, allowed the expansion of intact larval zebrafish, zebrafish embryos, Drosophila wing and mouse embryo (Sim et al., 2021, preprint; Steib et al., 2023; Steib et al., 2022). Overall, TissUExM was designed to ensure the isotropic fourfold expansion of heterogeneous tissue samples. TissUExM revealed endogenous proteins in the brain and spinal cord of whole zebrafish embryos to analyze cilia heterogeneity and tissue-specific defects associated with ciliopathies in multiple body locations simultaneously (Steib et al., 2022). These locations included the olfactory bulb and longitudinally-distributed structures such as the hair cells of the lateral line and cilia of the spinal cord. This approach permits the comparison of ultrastructural details in spatially distinct tissues within the same intact embryo.

Whole-body ExM (Figure 2) was successfully applied to larger, denser, and older animals, such as larval zebrafish up to 8 days post-fertilization and even juvenile fish (several months old). It enabled an increase in resolution from approximately 1 µm to 60 nm (Sim et al., 2021, preprint).

Both TissUExM and Whole-body ExM can achieve the expansion of an intact, several-millimeter-long animal and are compatible with simultaneous immunohistological staining and with genetically-encoded fluorescent protein labeling. Such techniques are conducive to the observation of intricate neural connections and signaling pathways throughout the nervous system from the brain to the spinal cord and allow for the exploration of various neural circuits involving complex spatially-distributed signals and connections that vary along the neuraxis. These ExM techniques provide an essential link between structure and function, revealing not only subsynaptic structures but also intricate neural connections and signaling pathways throughout the nervous system from the brain to the spinal cord.

**Unpacking synaptic connectivity in thick tissue of mice**

Mice (*Mus musculus*) are biomedical research's most widely used animal species. Because of the extensive literature on mice and well-established protocols for genetic manipulation, mice are used to address experimental questions on the relationship between genes and neural circuits and to test models of neurological and psychological diseases (Dawson et al., 2018). However, mice share the same resolution problem as other model systems where sub-cellular structures are below the diffraction limit. Furthermore, mouse brains are substantially larger than some non-mammalian species like zebrafish, and as a result, the brain has to be sectioned and reconstructed to allow light penetration for imaging. Imaging mouse tissue remains challenging, despite recent advances in tissue clearing (Chung et al., 2013; Renier et al., 2014). Fortunately, ExM also clears the sample in the digestion step, homogenizing the refractive index throughout the sample and limiting the scattering of photons within the sample (Chen et al., 2015). Below, we describe how anatomical regions deep in the mouse brain were studied with ExM and iterative RNA-FISH of brain slices. ExFISH combines molecular information, enabling the identification of cell-types and sub-regions, with the resolving of fine neural processes and structures (Wang et al., 2021). This advance was predicated upon the ability to combine RNA-FISH and ExM, pioneered as ExFISH (Chen et al., 2016). The RNA is stained by fluorescence in-situ hybridization (RNA-FISH) amplified with hybridization chain reactions (HCRs) that enable resolving single RNA molecules. Multiple rounds of FISH increase the number of target RNA sequences (Chen et al., 2016). In an elegant example of decrowding to understand nanostructure colocalization, Expansion Revealing (ExR) shows the distributed pathological nano-markers of Alzheimer's disease (Sarkar et al., 2022). These recent advances have opened new avenues of investigation to observe RNA and tightly clustered proteins deep in the mouse brain which were previously inaccessible. Magnify builds on the 10x and proExM protocols to enable nanoscale imaging across diverse biological samples, achieving resolutions as fine as 25 nm (Klimas et al., 2023). By preserving a broad range of biomolecules—including nucleic acids, proteins, and lipids—Magnify serves as the foundation for advanced protocols like ExIGS, a genomic DNA sequencing method (Labade et al., 2024).

. Expanding on these capabilities, methods like Expansion sequencing (ExSeq) and Expansion in situ genome sequencing (ExIGS) offer nanoscale resolution for in situ RNA and genome sequencing, respectively. ExSeq allows untargeted transcriptomics in complex tissues, such as tumor microenvironments and neuron-dense areas, while ExIGS links nuclear abnormalities to chromatin repression hotspots, illuminating how structural changes affect gene regulation, especially relevant in aging and disease research (Alon et al., 2021; Labade et al., 2024).

**Spatial transcriptomics with ExM**

Expansion Fluorescence *In Situ* Hybridization (ExFISH) (Figure 2), was the first ExM variant that introduced the anchoring of nucleic acids, such as RNA, to the gel (Chen et al., 2016). Expansion-Assisted Iterative Fluorescence *In Situ* Hybridization (EASI-FISH) (Figure 2), improves spatial precision of *in situ* hybridization in thick tissue specimens via protocol optimization and the use of a distinct anchoring molecule (Wang et al., 2021). EASI-FISH was used to spatially determine transcripts from many genes simultaneously across a 300 µm thick slice and identified previously unknown morphological diversity in deep regions such as the lateral hypothalamic area of mice. Using EASI-FISH in combination with genetic data from single-cell RNA-Sequencing (scRNA-Seq), researchers characterized marker genes in the lateral hypothalamic area and classified nine spatio-molecular regions (Wang et al., 2021).

**Revealing hidden protein structures with ExM**

In Expansion Revealing (ExR) (Figure 2), proteins are "decrowded" by two iterative expansion steps such that the increased distance between proteins allows post-expansion applied antibodies, larger than the space between these tightly packed proteins, to access their targets. Decrowding allows the labeling of targets which cannot be visualized well even using super-resolution techniques, since the targets are challenging to label precisely until they are expanded. Using ExR, researchers revealed the nanoscale coordination of presynaptic calcium channels with postsynaptic proteins and the periodic nanostructures of amyloid beta plaques in mouse models of Alzheimer's disease (Sarkar et al., 2022). This was only possible due to ExR's capacity to decrowd proteins spatially while imaging with resolution on par with super-resolution techniques.

## Looking across the entire brain with unparalleled resolution in *Drosophila*

The fruit fly (*Drosophila melanogaster*) is commonly used in neurobiological research due to its practical advantages, available genetic tools, and well-documented connectome (Scheffer et al., 2020). Flies are easy and inexpensive to maintain, have a short life cycle, and produce large numbers of offspring. *Drosophila* were first used to understand genetic inheritance, chromosome mapping, and genetic mutations (Morgan, 1909). They share many conserved genes with humans, including those involved in cell division, development, and basic body plan formation (Reiter et al., 2001). Approximately 75% of human disease genes have equivalents in flies (Reiter et al., 2001). Due to their highly tractable genetics and well-characterized behavioral repertoire, *Drosophila* can be used to identify neuronal circuits and genes underlying behavior (Devineni & Scaplen, 2021).

A complete synaptic-resolution connectome of the *Drosophila* larval brain (Winding et al., 2023), and an adult brain connectome (Dorkenwald et al., 2024) were published, based on datasets from electron microscopy. However impressive, projects on this scale with traditional imaging methods can be expensive and slow. These atlases require expensive and not broadly available equipment, significant computation time and resources, and often require large cohorts of research groups for their generation. In contrast, ExM can be applied within days in most equipped laboratories and makes it possible to collect high-resolution information on neural structure in multiple animals. High-throughput imaging of multiple animals makes it possible to address behavioral heterogeneity and their circuit origins: researchers can compare neural structures in cohorts of animals presenting distinct behaviors, tracking how neural structures and corresponding behaviors evolve during development or are modulated to mediate learning. When ExM was first applied to the imaging of *Drosophila* brains it was coupled with lattice-light sheet microscopy (LLSM) to obtain enhanced volumetric imaging resolution (Gao et al., 2019; Lillvis et al., 2022).

**In *Drosophila*, ExM enables the comparison of neural circuit connectivity characteristics between populations**

ExM was optimized for application to larval *Drosophila* brains. The presynaptic protein organization observed was consistent with that established with other super-resolution microscopy methods. Additionally, ExM allowed more precise quantification of insertion of somatosensory neural dendrites into epithelial cells compared to standard confocal microscopy (Jiang et al., 2018).

When expansion was followed by lattice light-sheet microscopy (ExLLSM, Gao et al., 2019) (Figure 2), it allowed for high-resolution imaging of the entire *Drosophila* brain, with the ability to distinguish features that are approximately 60 nm apart in the x and y dimensions, and 90 nm apart in the z dimension. Furthermore, ExLLSM image acquisition was approximately 700 times faster than other super-resolution fluorescent microscopy techniques and 1200 times faster than EM (Gao et al., 2019). The technique makes it possible to zoom in on individual synapses and zoom out to see broader patterns in synaptic connectivity, facilitating comprehensive and comparative anatomical investigations (Pesce et al., 2024).

Using ExLLSM, researchers imaged all dopaminergic neurons across the *Drosophila* brain, visualizing their morphologies in all major brain regions and tracing specific clusters to determine cell types (Gao et al., 2019). They also quantified presynaptic active zones across the brain, providing insights into the local density of synapses and dopaminergic neuron-associated active zones (Gao et al., 2019). ExLLSM has been used to trace axonal branches of olfactory projection neurons and study their bouton arrangements at the calyx and lateral horn across multiple *Drosophila* specimens (Gao et al., 2019).

Lillvis et al., 2022 added the development of a high-throughput ExLLSM open-source image analysis pipeline to enhance the accessibility of studying neuronal circuits. Importantly, in this manner they were able to produce a proof of concept for one of the greatest promises of the application of ExM to neural circuits: due to the high speed and applicability to volumetric samples, ExM can be applied for the comparison of neural circuit connectivity across animal populations. This is a major advance compared to EM connectome generation that provides higher resolution data but for a single animal. Thus, they demonstrated the quantification of synapses in male compared to female fly populations. In the future, this approach can be applied to reveal gender, development and experience-dependent differences in neural circuit and synapse structure. They also revealed synaptic structural underpinnings of behavior variability in a population. Such questions can only be addressed with rapid, volumetric imaging beyond the diffraction limit.

## **Localization of diverse molecules in *C. elegans***

The roundworm *C. elegans* is the first organism to have a complete connectome mapping synapses and gap junctions among neuron types (White et al., 1986). Its compact nervous system of 302 neurons is highly stereotyped and capable of encoding essential and interesting behaviors, ranging from reflexive escape from harsh touch (Chalfie & Sulston, 1981) to associative and nonassociative learning (Ardiel & Rankin, 2010). Additionally, because of its small size, with adults measuring around 1 mm long, optical transparency, and genetic tractability, the worm is well-suited for *in vivo* imaging of genetically-encoded fluorophores and sensors (Chalfie et al., 1994; Corsi et al., 2015).

The completeness of *C. elegans*'s community atlases and databases now spans connectomes of multiple developmental time points and of both sexes (Cook et al., 2019; Jarrell et al., 2012; White et al., 1986; Witvliet et al., 2021), cell lineage diagrams tracing the birth of every cell (Kimble & Hirsh, 1979; Sulston et al., 1983; Sulston & Horvitz, 1977), a pan-neuronal atlas strain labeling every neuron non-stochastically (Yemini et al., 2021), a signal propagation atlas outlining the functional partners of most head neurons (Randi et al., 2023), single-cell RNAseq transcriptomes (Ben-David et al., 2021; Hammarlund et al., 2018; Packer et al., 2019), and the first complete animal genome (*C. elegans* Sequencing Consortium, 1998; Yoshimura et al., 2019). This shared information lends itself to a highly collaborative field capable of high-resolution interrogation of fundamental neurobiological principles. However, many important subcellular structures are smaller than the diffraction limit of light microscopy. Electron Microscopy and, more recently, optical super-resolution have been the only methods to collect data about structures such as pre- and postsynaptic specializations, gap junctions, and distinct neurites in a bundle or neuropile.

## **Adapting Expansion Microscopy for *C. elegans***

Super-resolution microscopy methods such as STORM, PALM, SR-SIM, or STED have been used to study intact and dissected C. elegans, (Gao et al., 2012; He et al., 2016; Köhler et al., 2017; Krieg et al., 2017; Rankin et al., 2011; Vangindertael et al., 2015) but the imaging depth of these techniques is insufficient to map the entire depth of the animal. Furthermore, the tough cuticle of C. elegans limits antibody penetration and immunohistochemistry important for STORM or STED. The technical difficulty of immunostaining in C. elegans likely contributed to the pioneering use of GFP as a reporter in the worm. However, in recent years, the first ExM protocol for C. elegans has introduced a cuticle permeabilization technique for immunostaining and expansion (Yu et al., 2020) (Figure 2). Immunostaining allows robust labeling of structures of interest for high-resolution imaging of a large region of interest with reduced constraints from photobleaching of endogenously produced fluorophores (Yu et al., 2020; Yu et al., 2022). In this initial proof-of-concept paper, the authors show that ExM can resolve adjacent synaptic puncta, more than doubling the number of puncta counted pre-expansion (Yu et al., 2020).

Several technical challenges remain for a comprehensive optical connectome (Sneve & Piatkevich, 2022). The first challenge is in the complete and continuous segmentation of each neuron, where the identity of each neuron can be maintained across the reach of its neurites. A novel probe has recently been developed expressly for this purpose of dense and continuous membrane labeling in ExM; this azide probe binds to fixed tissue membranes and later is fluorescently labeled via click-chemistry (Shin et al., 2024, preprint). The next challenge is in labeling pre- and postsynaptic specializations while retaining contextual information about the neuronal identities. Given the mere 20 nanometer space between opposed membranes of synapses (Palay, 1956) and the enormous synaptic density, conventional light microscopy cannot parse fluorophore-labeled synapses with the resolution and throughput required to build a connectome. However, with up to 20x physical magnification of samples, ExM could disambiguate individual GFP molecules in the cytosol of neurons (Yu et al., 2020); such resolution has the potential to unpack dense neuronal architecture for optical connectomics.

In the way of protocol optimization and proof of concept, Yu et al. (2020, 2022) demonstrated two relevant neuronal features that could be resolved by ExM. One is the disambiguation of individual neurites within a bundle and through defasciculation, as tracking neurites and maintaining neuronal identity is crucial when collecting connectome data. The other feature is the localization of chemical synapses and gap junctions by immunostaining and fluorescence *in situ*. While synaptic partners can be deduced from membrane proximity, assigning electric and chemical connections, the directions of synaptic connections, and even neurotransmitters, depending on staining specificity, can greatly enrich the connectome data (Brittin et al., 2021). These qualitative demonstrations were not driven by an explicit hypothesis but suggest how ExM can provide connectome data.

Thus far, quantitative studies using ExM in *C. elegans* have been mostly cellular rather than neuronal, for example, the ultrastructure organization of P-granules and the dynamics of centrioles. ExM revealed the spatial arrangement of proteins that bind small RNAs in the oocyte P-granule (Suen et al., 2023). P-granule are membraneless organelles that are detectable by EM, but their protein ultrastructures cannot be discerned because the proteins cannot be differentially labeled for EM. During centriole elimination in oocytes, ExM has been used to reveal the sequence of ultra-structural changes with high spatiotemporal fidelity, where prior analyses have been limited to serial-section EM or immunohistochemistry for either high spatial or temporal resolution, respectively (Pierron et al., 2023). Building upon recent advances in the bio-orthogonal click chemistry, multifunctional ExM linkers have been developed to link and label glycans in the ExM gel with high specificity and isotropy, visualizing the distribution of glycans in anatomical context with nanoscale resolution (Kuo et al., 2024, preprint). Applying click chemistry and its compatibility with ExM adds versatility to label almost limitless molecular targets.

**Methods for Expansion in *C. elegans* (ExCEL)**

Expansion in *C. elegans* (ExCel) has primarily been developed by the Boyden laboratory at MIT (Yu et al., 2020; Yu et al., 2022). There are three published protocols for magnifying fixed, whole animals: a standard, an epitope-preserving, and an iterative version (Table 1). The standard protocol allows visualization of fluorescent proteins, which are stable enough to withstand Proteinase K digestion. The epitope-preserving protocol uses a gentler digest that allows most endogenous proteins to be labeled by off-the-shelf antibodies. Iterative expansion allows for the highest spatial resolution (~25 nm). Imaging on a standard confocal microscope provides the following resolutions and limitations.

**Table 1. Summary of Expansion Microscopy techniques in *C. elegans*.** Table constructed using information from (Yu et al., 2022).

| ExCel variation | Standard ExCEL | Epitope-preserving ExCEL | Iterative ExCEL |
| --- | --- | --- | --- |
| Molecular readout | Fluorescent proteins, RNA, DNA, general anatomy | Endogenous proteins | Fluorescent proteins |
| Linear Expansion | 3.5x linear expansion<br>70 nm resolution<br>High isotropy (1-6% error) | 2.8x linear expansion<br>100 nm resolution<br>Moderate isotropy (8-25% error) | 20x linear expansion<br>25 nm resolution<br>High isotropy (1.5-4.5% error) |
| Protocol duration | 10 days for fluorophores<br>14 days for fluorophores and RNAs<br>+1 day for amine-stain for anatomy | 18 days | 19 days |
| Limitations | RNA, DNA, and anatomy can be revealed through stains or *in situ* hybridization.<br><br>Proteins of interest must be labeled with fluorophores, which are structurally able to withstand the Proteinase K digestion that softens tissue for expansion and permeabilizes the worm cuticle for antibody access. | Proteinase K is replaced with a gentler cuticle-disrupting collagenase and heat-mediated protein digestion. This leads to slightly worse expansion isotropy but allows for readout of ~70% of endogenous proteins. | Following Proteinase K digestion, an oligonucleotide-conjugated antibody is applied to stain target fluorophores. The oligonucleotide transfers and amplifies stained signal across two rounds of tissue expansion.<br><br>Iterative ExCel is more technically demanding and yields weaker fluorescent signal per voxel than standard ExCel. |

**Summary**

ExM, in its many varieties, has been used to visualize previously unresolvable subcellular and nano-structures associated with synapses and neuronal processes, in addition to nucleic acids. In the context of neural circuit studies, it was shown to be applicable to resolving fine neural processes, putative synaptic connections between neural populations, fine intra-synaptic structures in both the pre- and post- synaptic specializations, and interactions between glia and neurons. Expansion can now be performed up to twentyfold and be coupled to imaging with a broad variety of fluorescent microscopy technologies, including sub-diffraction microscopy methods, and custom protocols have been published for many organisms. With an appropriate selection of: 1) the staining method, including the decision of whether to stain before expansion or use the protein "decrowding" effect of expansion, which makes targets more accessible and staining more accurate; 2) of the expansion protocol; and 3) of the post-expansion imaging method, it becomes possible to capture the smallest neural features while imaging large neural circuits and adapt the imaging abilities to the requirements of the scientific question being addressed.

The ease of adaptation without the need to obtain new and expensive equipment or alter established staining protocols, the relative simplicity of the protocols, and the unprecedented ability to apply nano-scale imaging to large tissue volumes at high throughput represent a major leap forward towards enabling significant advances in our depth of knowledge about the nervous system. Additional important advances include the automation of sample preparation, imaging, and analysis, facilitating the application of nano-scale imaging and analysis to large animal cohorts.

With a nano-scale imaging technology applicable at high throughput to any genetic model organism in which neural circuit studies are conducted, ExM represents a unique and powerful method for addressing outstanding questions regarding neural circuit structure and function relation. With ExM, we can look at how connectivity patterns and intra-synaptic structure are co-modulated during learning or development to give rise to altered relations between sensory inputs and motor outputs, or unravel the structural underpinnings of functional heterogeneity in the performance of sensory-motor transformations and in their adaptation by experience.

**ACKNOWLEDGEMENTS**

We thank Dr. Eric Fortune for valuable critical comments on the manuscript. Gal Haspel is funded by the National Institute of Neurological Disorders and Stroke (1R15NS125565-01); and the US Department of Energy Office of Science, Office of Fusion Energy Sciences (DE-SC0023430). Limor Freifeld was funded by the Chief Scientist Office (CSO) of the Israeli Ministry of Health (MOH) as part of an ERA-NET co-fund scheme and by the Zuckerman STEM Leadership Program.

**Conflict of Interest Statement:** The authors declare no conflict of interest.

**Code and Data Availability Statement:** The authors have generated no code or data in this review.

**REFERENCES**

Abbe, E. (1873). Beiträge zur Theorie des Mikroskops und der mikroskopischen Wahrnehmung. *Archiv für Mikroskopische Anatomie*, *9*(1), 413–468. https://doi.org/10.1007/BF02956173

Ahrens, M. B., Li, J. M., Orger, M. B., Robson, D. N., Schier, A. F., Engert, F., & Portugues, R. (2012). Brain-wide neuronal dynamics during motor adaptation in zebrafish. *Nature*, *485*(7399), 471–477. https://doi.org/10.1038/nature11057

Almada, P., Culley, S., & Henriques, R. (2015). PALM and STORM: Into large fields and high-throughput microscopy with sCMOS detectors. *Methods (San Diego, Calif.)*, *88*, 109–121. https://doi.org/10.1016/j.ymeth.2015.06.004

Alon, S., Goodwin, D. R., Sinha, A., Wassie, A. T., Chen, F., Daugharthy, E. R., Bando, Y., Kajita, A., Xue, A. G., Marrett, K., Prior, R., Cui, Y., Payne, A. C., Yao, C.-C., Suk, H.-J., Wang, R., Yu, C.-C. (Jay), Tillberg, P., Reginato, P., … Boyden, E. S. (2021). Expansion sequencing: Spatially precise in situ transcriptomics in intact biological systems. Science, 371(6528), eaax2656. https://doi.org/10.1126/science.aax2656

Alvarez, F. J., Dewey, D. E., Harrington, D. A., & Fyffe, R. E. (1997). Cell-type specific organization of glycine receptor clusters in the mammalian spinal cord. *The Journal of Comparative Neurology*, *379*(1), 150–170.

Antinucci, P., & Hindges, R. (2016). A crystal-clear zebrafish for in vivo imaging. *Scientific Reports*, *6*(1), 29490. https://doi.org/10.1038/srep29490

Arizono, M., Idziak, A., Quici, F., & Nägerl, U. V. (2023). Getting sharper: The brain under the spotlight of super-resolution microscopy. *Trends in Cell Biology*, *33*(2), 148–161. https://doi.org/10.1016/j.tcb.2022.06.011

Asano, S. M., Gao, R., Wassie, A. T., Tillberg, P. W., Chen, F., & Boyden, E. S. (2018). Expansion Microscopy: Protocols for Imaging Proteins and RNA in Cells and Tissues. *Current Protocols in Cell Biology*, *80*(1), e56. https://doi.org/10.1002/cpcb.56

Ben-David, E., Boocock, J., Guo, L., Zdraljevic, S., Bloom, J. S., & Kruglyak, L. (2021). Whole-organism eQTL mapping at cellular resolution with single-cell sequencing. *eLife*, *10*, e65857. https://doi.org/10.7554/eLife.65857

Betzig, E., Patterson, G. H., Sougrat, R., Lindwasser, O. W., Olenych, S., Bonifacino, J. S., Davidson, M. W., Lippincott-Schwartz, J., & Hess, H. F. (2006). Imaging intracellular fluorescent proteins at nanometer resolution. *Science (New York, N.Y.)*, *313*(5793), 1642–1645. https://doi.org/10.1126/science.1127344

Betzig, E., & Trautman, J. K. (1992). Near-Field Optics: Microscopy, Spectroscopy, and Surface Modification Beyond the Diffraction Limit. *Science*, *257*(5067), 189–195. https://doi.org/10.1126/science.257.5067.189

Brittin, C. A., Cook, S. J., Hall, D. H., Emmons, S. W., & Cohen, N. (2021). A multi-scale brain map derived from whole-brain volumetric reconstructions. *Nature*, *591*(7848), 105–110. https://doi.org/10.1038/s41586-021-03284-x

C. elegans Sequencing Consortium. (1998). Genome sequence of the nematode C. elegans: A platform for investigating biology. *Science (New York, N.Y.)*, *282*(5396), 2012–2018. https://doi.org/10.1126/science.282.5396.2012

Campbell, B. C., Nabel, E. M., Murdock, M. H., Lao-Peregrin, C., Tsoulfas, P., Blackmore, M. G., Lee, F. S., Liston, C., Morishita, H., & Petsko, G. A. (2020). mGreenLantern: A bright monomeric fluorescent protein with rapid expression and cell filling properties for neuronal imaging. *Proceedings of the National Academy of Sciences of the United States of America*, *117*(48), 30710–30721. https://doi.org/10.1073/pnas.2000942117

Sandra P Cárdenas-García, Sundas Ijaz, Alberto E Pereda (2024). The components of an electrical synapse as revealed by expansion microscopy of a single synaptic contact. *eLife*, 13:e91931. https://doi.org/10.7554/eLife.91931

Chalfie, M., & Sulston, J. (1981). Developmental genetics of the mechanosensory neurons of Caenorhabditis elegans. *Developmental Biology*, *82*(2), 358–370. https://doi.org/10.1016/0012-1606(81)90459-0

Chalfie, M., Tu, Y., Euskirchen, G., Ward, W. W., & Prasher, D. C. (1994). Green fluorescent protein as a marker for gene expression. *Science (New York, N.Y.)*, *263*(5148), 802–805. https://doi.org/10.1126/science.8303295

Chang, J.-B., Chen, F., Yoon, Y.-G., Jung, E. E., Babcock, H., Kang, J. S., Asano, S., Suk, H.-J., Pak, N., Tillberg, P. W., Wassie, A. T., Cai, D., & Boyden, E. S. (2017). Iterative expansion microscopy. *Nature Methods*, *14*(6), 593–599. https://doi.org/10.1038/nmeth.4261

Chen, F., Tillberg, P. W., & Boyden, E. S. (2015). Expansion microscopy. *Science*, *347*(6221), 543–548. https://doi.org/10.1126/science.1260088

Chen, F., Wassie, A. T., Cote, A. J., Sinha, A., Alon, S., Asano, S., Daugharthy, E. R., Chang, J.-B., Marblestone, A., Church, G. M., Raj, A., & Boyden, E. S. (2016). Nanoscale imaging of RNA with expansion microscopy. *Nature Methods*, *13*(8), 679–684. https://doi.org/10.1038/nmeth.3899

Chung, K., Wallace, J., Kim, S.-Y., Kalyanasundaram, S., Andalman, A. S., Davidson, T. J., Mirzabekov, J. J., Zalocusky, K. A., Mattis, J., Denisin, A. K., Pak, S., Bernstein, H., Ramakrishnan, C., Grosenick, L., Gradinaru, V., & Deisseroth, K. (2013). Structural and molecular interrogation of intact biological systems. *Nature*, *497*(7449), 332–337. https://doi.org/10.1038/nature12107

Cook, S. J., Jarrell, T. A., Brittin, C. A., Wang, Y., Bloniarz, A. E., Yakovlev, M. A., Nguyen, K. C. Q., Tang, L. T.-H., Bayer, E. A., Duerr, J. S., Bülow, H. E., Hobert, O., Hall, D. H., & Emmons, S. W. (2019). Whole-animal connectomes of both Caenorhabditis elegans sexes. *Nature*, *571*(7763), 63–71. https://doi.org/10.1038/s41586-019-1352-7

Corsi, A. K., Wightman, B., & Chalfie, M. (2015). A Transparent Window into Biology: A Primer on Caenorhabditis elegans. *Genetics*, *200*(2), 387–407. https://doi.org/10.1534/genetics.115.176099

Damstra, H. G., Mohar, B., Eddison, M., Akhmanova, A., Kapitein, L. C., & Tillberg, P. W. (2022). Visualizing cellular and tissue ultrastructure using Ten-fold Robust Expansion Microscopy (TREx). *eLife*, *11*, e73775. https://doi.org/10.7554/eLife.73775

Dawson, T. M., Golde, T. E., & Lagier-Tourenne, C. (2018). Animal models of neurodegenerative diseases. *Nature Neuroscience*, *21*(10), 1370–1379. https://doi.org/10.1038/s41593-018-0236-8

Dekker, A. J. den, & Bos, A. van den. (1997). Resolution: A survey. *JOSA A*, *14*(3), 547–557. https://doi.org/10.1364/JOSAA.14.000547

Devineni, A. V., & Scaplen, K. M. (2021). Neural Circuits Underlying Behavioral Flexibility: Insights From Drosophila. *Frontiers in Behavioral Neuroscience*, *15*, 821680. https://doi.org/10.3389/fnbeh.2021.821680

Dorkenwald, S., Matsliah, A., Sterling, A. R., Schlegel, P., Yu, S., McKellar, C. E., Lin, A., Costa, M., Eichler, K., Yin, Y., Silversmith, W., Schneider-Mizell, C., Jordan, C. S., Brittain, D., Halageri, A., Kuehner, K., Ogedengbe, O., Morey, R., Gager, J., … Murthy, M. (2024). *Neuronal wiring diagram of an adult brain. Nature, 634(8032), 124–138*. https://doi.org/10.1038/s41586-024-07558-y

Eaton, R. C., Lee, R. K. K., & Foreman, M. B. (2001). The Mauthner cell and other identified neurons of the brainstem escape network of fish. *Progress in Neurobiology*, *63*(4), 467–485. https://doi.org/10.1016/S0301-0082(00)00047-2

Freifeld, L., Odstrcil, I., Förster, D., Ramirez, A., Gagnon, J. A., Randlett, O., Costa, E. K., Asano, S., Celiker, O. T., Gao, R., Martin-Alarcon, D. A., Reginato, P., Dick, C., Chen, L., Schoppik, D., Engert, F., Baier, H., & Boyden, E. S. (2017). Expansion microscopy of zebrafish for neuroscience and developmental biology studies. *Proceedings of the National Academy of Sciences*, *114*(50), E10799–E10808. https://doi.org/10.1073/pnas.1706281114

Gambarotto, D., Zwettler, F. U., Le Guennec, M., Schmidt-Cernohorska, M., Fortun, D., Borgers, S., Heine, J., Schloetel, J.-G., Reuss, M., Unser, M., Boyden, E. S., Sauer, M., Hamel, V., & Guichard, P. (2019). Imaging cellular ultrastructures using expansion microscopy (U-ExM). *Nature Methods*, *16*(1), 71–74. https://doi.org/10.1038/s41592-018-0238-1

Gao, L., Shao, L., Higgins, C. D., Poulton, J. S., Peifer, M., Davidson, M. W., Wu, X., Goldstein, B., & Betzig, E. (2012). Noninvasive Imaging beyond the Diffraction Limit of 3D Dynamics in Thickly Fluorescent Specimens. *Cell*, *151*(6), 1370–1385. https://doi.org/10.1016/j.cell.2012.10.008

Gao, R., Asano, S. M., Upadhyayula, S., Pisarev, I., Milkie, D. E., Liu, T.-L., Singh, V., Graves, A., Huynh, G. H., Zhao, Y., Bogovic, J., Colonell, J., Ott, C. M., Zugates, C., Tappan, S., Rodriguez, A., Mosaliganti, K. R., Sheu, S.-H., Pasolli, H. A., … Betzig, E. (2019). Cortical column and whole-brain imaging with molecular contrast and nanoscale resolution. *Science (New York, N.Y.)*, *363*(6424), eaau8302. https://doi.org/10.1126/science.aau8302

Gao, R., Yu, C.-C. (Jay), Gao, L., Piatkevich, K. D., Neve, R. L., Munro, J. B., Upadhyayula, S., & Boyden, E. S. (2021). A highly homogeneous polymer composed of tetrahedron-like monomers for high-isotropy expansion microscopy. *Nature Nanotechnology*, *16*(6), 698–707. https://doi.org/10.1038/s41565-021-00875-7

Gustafsson, M. G. (2000). Surpassing the lateral resolution limit by a factor of two using structured illumination microscopy. *Journal of Microscopy*, *198*(Pt 2), 82–87. https://doi.org/10.1046/j.1365-2818.2000.00710.x

Hale, M. E., Katz, H. R., Peek, M. Y., & Fremont, R. T. (2016). Neural circuits that drive startle behavior, with a focus on the Mauthner cells and spiral fiber neurons of fishes. *Journal of Neurogenetics*, *30*(2), 89–100. https://doi.org/10.1080/01677063.2016.1182526

Hamel, V., & Guichard, P. (2021). Improving the resolution of fluorescence nanoscopy using post-expansion labeling microscopy. *Methods in Cell Biology*, *161*, 297–315. https://doi.org/10.1016/bs.mcb.2020.07.002

Hammarlund, M., Hobert, O., Miller, D. M., & Sestan, N. (2018). The CeNGEN Project: The Complete Gene Expression Map of an Entire Nervous System. *Neuron*, *99*(3), 430–433. https://doi.org/10.1016/j.neuron.2018.07.042

He, J., Zhou, R., Wu, Z., Carrasco, M. A., Kurshan, P. T., Farley, J. E., Simon, D. J., Wang, G., Han, B., Hao, J., Heller, E., Freeman, M. R., Shen, K., Maniatis, T., Tessier-Lavigne, M., & Zhuang, X. (2016). Prevalent presence of periodic actin–spectrin-based membrane skeleton in a broad range of neuronal cell types and animal species. *Proceedings of the National Academy of Sciences*, *113*(21), 6029–6034. https://doi.org/10.1073/pnas.1605707113

Hell, S. W., & Wichmann, J. (1994). Breaking the diffraction resolution limit by stimulated emission: Stimulated-emission-depletion fluorescence microscopy. *Optics Letters*, *19*(11), 780–782. https://doi.org/10.1364/OL.19.000780

Higashijima, S., Masino, M. A., Mandel, G., & Fetcho, J. R. (2003). Imaging neuronal activity during zebrafish behavior with a genetically encoded calcium indicator. *Journal of Neurophysiology*, *90*(6), 3986–3997. https://doi.org/10.1152/jn.00576.2003

Hugelier, S., Colosi, P. L., & Lakadamyali, M. (2023). Quantitative Single-Molecule Localization Microscopy. *Annual Review of Biophysics*, *52*(Volume 52, 2023), 139–160. https://doi.org/10.1146/annurev-biophys-111622-091212

Hümpfer, N., Thielhorn, R., & Ewers, H. (2024). Expanding boundaries – a cell biologist's guide to expansion microscopy. *Journal of Cell Science*, *137*(7), jcs260765. https://doi.org/10.1242/jcs.260765

Huszka, G., & Gijs, M. A. M. (2019). Super-resolution optical imaging: A comparison. *Micro and Nano Engineering*, *2*, 7–28. https://doi.org/10.1016/j.mne.2018.11.005

Jarrell, T. A., Wang, Y., Bloniarz, A. E., Brittin, C. A., Xu, M., Thomson, J. N., Albertson, D. G., Hall, D. H., & Emmons, S. W. (2012). The Connectome of a Decision-Making Neural Network. *Science*, *337*(6093), 437–444. https://doi.org/10.1126/science.1221762

Jiang, N., Kim, H.-J., Chozinski, T. J., Azpurua, J. E., Eaton, B. A., Vaughan, J. C., & Parrish, J. Z. (2018). Superresolution imaging of Drosophila tissues using expansion microscopy. *Molecular Biology of the Cell*, *29*(12), 1413–1421. https://doi.org/10.1091/mbc.E17-10-0583

Kim, D. H., Kim, J., Marques, J. C., Grama, A., Hildebrand, D. G. C., Gu, W., Li, J. M., & Robson, D. N. (2017). Pan-neuronal calcium imaging with cellular resolution in freely swimming zebrafish. *Nature Methods*, *14*(11), 1107–1114. https://doi.org/10.1038/nmeth.4429

Kimble, J., & Hirsh, D. (1979). The postembryonic cell lineages of the hermaphrodite and male gonads in Caenorhabditis elegans. *Developmental Biology*, *70*(2), 396–417. https://doi.org/10.1016/0012-1606(79)90035-6

Klimas, A., Gallagher, B. R., Wijesekara, P., Fekir, S., DiBernardo, E. F., Cheng, Z., Stolz, D. B., Cambi, F., Watkins, S. C., Brody, S. L., Horani, A., Barth, A. L., Moore, C. I., Ren, X., & Zhao, Y. (2023). Magnify is a universal molecular anchoring strategy for expansion microscopy. *Nature Biotechnology*, *41*(6), 858–869. https://doi.org/10.1038/s41587-022-01546-1

Klimas, A., Gallagher, B., & Zhao, Y. (2019). The Basics of Expansion Microscopy. *Current Protocols in Cytometry*, *91*(1), e67. https://doi.org/10.1002/cpcy.67

Köhler, S., Wojcik, M., Xu, K., & Dernburg, A. F. (2017). Superresolution microscopy reveals the three-dimensional organization of meiotic chromosome axes in intact *Caenorhabditis elegans* tissue. *Proceedings of the National Academy of Sciences*, *114*(24). https://doi.org/10.1073/pnas.1702312114

Koyama, M., Kinkhabwala, A., Satou, C., Higashijima, S., & Fetcho, J. (2011). Mapping a sensory-motor network onto a structural and functional ground plan in the hindbrain. *Proceedings of the National Academy of Sciences of the United States of America*, *108*(3), 1170–1175. https://doi.org/10.1073/pnas.1012189108

Krieg, M., Stühmer, J., Cueva, J. G., Fetter, R., Spilker, K., Cremers, D., Shen, K., Dunn, A. R., & Goodman, M. B. (2017). Genetic defects in β-spectrin and tau sensitize C. elegans axons to movement-induced damage via torque-tension coupling. *eLife*, *6*, e20172. https://doi.org/10.7554/eLife.20172

Kuo, J. C.-H., Colville, M. J., Sorkin, M. R., Kuo, J. L. K., Huang, L. T., Thornlow, D. N., Beacham, G. M., Hollopeter, G., DeLisa, M. P., Alabi, C. A., & Paszek, M. J. (2024). *Bio-orthogonal Glycan Imaging of Culture Cells and Whole Animal* C. elegans *with Expansion Microscopy*. https://doi.org/10.1101/2024.02.01.578333

Labade, A. S., Chiang, Z. D., Comenho, C., Reginato, P. L., Payne, A. C., Earl, A. S., Shrestha, R., Duarte, F. M., Habibi, E., Zhang, R., Church, G. M., Boyden, E. S., Chen, F., & Buenrostro, J. D. (2024). Expansion in situ genome sequencing links nuclear abnormalities to hotspots of aberrant euchromatin repression (p. 2024.09.24.614614). bioRxiv. https://doi.org/10.1101/2024.09.24.614614

Lauterbach, M. A., Ronzitti, E., Sternberg, J. R., Wyart, C., & Emiliani, V. (2015). Fast Calcium Imaging with Optical Sectioning via HiLo Microscopy. *PLOS ONE*, *10*(12), e0143681. https://doi.org/10.1371/journal.pone.0143681

Lillvis, J. L., Otsuna, H., Ding, X., Pisarev, I., Kawase, T., Colonell, J., Rokicki, K., Goina, C., Gao, R., Hu, A., Wang, K., Bogovic, J., Milkie, D. E., Meienberg, L., Mensh, B. D., Boyden, E. S., Saalfeld, S., Tillberg, P. W., & Dickson, B. J. (2022). Rapid reconstruction of neural circuits using tissue expansion and light sheet microscopy. *eLife*, *11*, e81248. https://doi.org/10.7554/eLife.81248

Louvel, V., Haase, R., Mercey, O., Laporte, M. H., Eloy, T., Baudrier, É., Fortun, D., Soldati-Favre, D., Hamel, V., & Guichard, P. (2023). iU-ExM: Nanoscopy of organelles and tissues with iterative ultrastructure expansion microscopy. *Nature Communications*, *14*(1), 7893. https://doi.org/10.1038/s41467-023-43582-8

Migault, G., van der Plas, T. L., Trentesaux, H., Panier, T., Candelier, R., Proville, R., Englitz, B., Debrégeas, G., & Bormuth, V. (2018). Whole-Brain Calcium Imaging during Physiological Vestibular Stimulation in Larval Zebrafish. *Current Biology*, *28*(23), 3723-3735.e6. https://doi.org/10.1016/j.cub.2018.10.017

Morgan, T. H. (1909). What are "Factors" in Mendelian Explanations? *Journal of Heredity*, *os-5*(1), 365–367. https://doi.org/10.1093/jhered/os-5.1.365

Mu, Y., Bennett, D. V., Rubinov, M., Narayan, S., Yang, C.-T., Tanimoto, M., Mensh, B. D., Looger, L. L., & Ahrens, M. B. (2019). Glia Accumulate Evidence that Actions Are Futile and Suppress Unsuccessful Behavior. *Cell*, *178*(1), 27-43.e19. https://doi.org/10.1016/j.cell.2019.05.050

Packer, J. S., Zhu, Q., Huynh, C., Sivaramakrishnan, P., Preston, E., Dueck, H., Stefanik, D., Tan, K., Trapnell, C., Kim, J., Waterston, R. H., & Murray, J. I. (2019). A lineage-resolved molecular atlas of C. elegans embryogenesis at single-cell resolution. *Science (New York, N.Y.)*, *365*(6459), eaax1971. https://doi.org/10.1126/science.aax1971

Palay, S. L. (1956). Synapses in the central nervous system. *The Journal of Biophysical and Biochemical Cytology*, *2*(4 Suppl), 193–202. https://doi.org/10.1083/jcb.2.4.193

Perelsman, O., Asano, S., & Freifeld, L. (2022). Expansion Microscopy of Larval Zebrafish Brains and Zebrafish Embryos. In B. Heit (Ed.), *Fluorescent Microscopy* (pp. 211–222). Springer US. https://doi.org/10.1007/978-1-0716-2051-9_13

Pesce, L., Ricci, P., Sportelli, G., Belcari, N., & Sancataldo, G. (2024). Expansion and Light-Sheet Microscopy for Nanoscale 3D Imaging. *Small Methods*, *n/a*(n/a), 2301715. https://doi.org/10.1002/smtd.202301715

Piatkevich, K. D., & Boyden, E. S. (2024). Optogenetic control of neural activity: The biophysics of microbial rhodopsins in neuroscience. *Quarterly Reviews of Biophysics*, *57*, e1. https://doi.org/10.1017/S0033583523000033

Pierron, M., Woglar, A., Busso, C., Jha, K., Mikeladze-Dvali, T., Croisier, M., & Gönczy, P. (2023). Centriole elimination during Caenorhabditis elegans oogenesis initiates with loss of the central tube protein SAS-1. *The EMBO Journal*, *42*(24), e115076. https://doi.org/10.15252/embj.2023115076

Portugues, R., Feierstein, C. E., Engert, F., & Orger, M. B. (2014). Whole-brain activity maps reveal stereotyped, distributed networks for visuomotor behavior. *Neuron*, *81*(6), 1328–1343. https://doi.org/10.1016/j.neuron.2014.01.019

Randi, F., Sharma, A. K., Dvali, S., & Leifer, A. M. (2023). Neural signal propagation atlas of Caenorhabditis elegans. *Nature*, *623*(7986), 406–414. https://doi.org/10.1038/s41586-023-06683-4

Rankin, B. R., Moneron, G., Wurm, C. A., Nelson, J. C., Walter, A., Schwarzer, D., Schroeder, J., Colón-Ramos, D. A., & Hell, S. W. (2011). Nanoscopy in a Living Multicellular Organism Expressing GFP. *Biophysical Journal*, *100*(12), L63–L65. https://doi.org/10.1016/j.bpj.2011.05.020

Reiter, L. T., Potocki, L., Chien, S., Gribskov, M., & Bier, E. (2001). A Systematic Analysis of Human Disease-Associated Gene Sequences In Drosophila melanogaster. *Genome Research*, *11*(6), 1114–1125. https://doi.org/10.1101/gr.169101

Renier, N., Wu, Z., Simon, D. J., Yang, J., Ariel, P., & Tessier-Lavigne, M. (2014). iDISCO: A simple, rapid method to immunolabel large tissue samples for volume imaging. *Cell*, *159*(4), 896–910. https://doi.org/10.1016/j.cell.2014.10.010

Renninger, S. L., & Orger, M. B. (2013). Two-photon imaging of neural population activity in zebrafish. *Methods*, *62*(3), 255–267. https://doi.org/10.1016/j.ymeth.2013.05.016

Rust, M. J., Bates, M., & Zhuang, X. (2006). Sub-diffraction-limit imaging by stochastic optical reconstruction microscopy (STORM). *Nature Methods*, *3*(10), 793–796. https://doi.org/10.1038/nmeth929

Sarkar, D., Kang, J., Wassie, A. T., Schroeder, M. E., Peng, Z., Tarr, T. B., Tang, A.-H., Niederst, E. D., Young, J. Z., Su, H., Park, D., Yin, P., Tsai, L.-H., Blanpied, T. A., & Boyden, E. S. (2022). Revealing nanostructures in brain tissue via protein decrowding by iterative expansion microscopy. *Nature Biomedical Engineering*, *6*(9), 1057–1073. https://doi.org/10.1038/s41551-022-00912-3

Scheffer, L. K., Xu, C. S., Januszewski, M., Lu, Z., Takemura, S., Hayworth, K. J., Huang, G. B., Shinomiya, K., Maitlin-Shepard, J., Berg, S., Clements, J., Hubbard, P. M., Katz, W. T., Umayam, L., Zhao, T., Ackerman, D., Blakely, T., Bogovic, J., Dolafi, T., … Plaza, S. M. (2020). A connectome and analysis of the adult Drosophila central brain. *eLife*, *9*, e57443. https://doi.org/10.7554/eLife.57443

Seitanidou, T., Triller, A., & Korn, H. (1988). Distribution of glycine receptors on the membrane of a central neuron: An immunoelectron microscopy study. *Journal of Neuroscience*, *8*(11), 4319–4333. https://doi.org/10.1523/JNEUROSCI.08-11-04319.1988

Shi, X., Li, Q., Dai, Z., Tran, A. A., Feng, S., Ramirez, A. D., Lin, Z., Wang, X., Chow, T. T., Chen, J., Kumar, D., McColloch, A. R., Reiter, J. F., Huang, E. J., Seiple, I. B., & Huang, B. (2021). Label-retention expansion microscopy. *The Journal of Cell Biology*, *220*(9), e202105067. https://doi.org/10.1083/jcb.202105067

Shin, T. W.; Wang, H.; Zhang, C.; An, B.; Lu, Y.; Zhang, E.; Lu, X.; Karagiannis, E. D.; Kang, J. S.; Emenari, A.; Symvoulidis, P.; Asano, S.; Lin, L.; Costa, E. K.; Marblestone, A. H.; Kasthuri, N.; Tsai, L.-H.; Boyden, E. S. (2024). Dense, Continuous Membrane Labeling and Expansion Microscopy Visualization of Ultrastructure in Tissues. *bioRxiv*, 2024.03.07.583776. https://doi.org/10.1101/2024.03.07.583776.

Sim, J., Park, C. E., Cho, I., Min, K., Eom, M., Han, S., Jeon, H., Cho, H.-J., Cho, E.-S., Kumar, A., Chong, Y., Kang, J. S., Piatkevich, K. D., Jung, E. E., Kang, D.-S., Kwon, S.-K., Kim, J., Yoon, K.-J., Lee, J.-S., … Chang, J.-B. (2022). *Nanoscale resolution imaging of the whole mouse embryos and larval zebrafish using expansion microscopy* (p. 2021.05.18.443629). bioRxiv. https://doi.org/10.1101/2021.05.18.443629

Simmich, J., Staykov, E., & Scott, E. (2012). Zebrafish as an appealing model for optogenetic studies. *Progress in Brain Research*, *196*, 145–162. https://doi.org/10.1016/B978-0-444-59426-6.00008-2

Sneve, M. A., & Piatkevich, K. D. (2022). Towards a Comprehensive Optical Connectome at Single Synapse Resolution via Expansion Microscopy. *Frontiers in Synaptic Neuroscience*, *13*, 754814. https://doi.org/10.3389/fnsyn.2021.754814

Steib, E., Tetley, R., Laine, R. F., Norris, D. P., Mao, Y., & Vermot, J. (2022). TissUExM enables quantitative ultrastructural analysis in whole vertebrate embryos by expansion microscopy. *Cell Reports Methods*, *2*(10), 100311. https://doi.org/10.1016/j.crmeth.2022.100311

Steib, E., Vagena-Pantoula, C., & Vermot, J. (2023). TissUExM protocol for ultrastructure expansion microscopy of zebrafish larvae and mouse embryos. *STAR Protocols*, *4*(2), 102257. https://doi.org/10.1016/j.xpro.2023.102257

Sulston, J. E., & Horvitz, H. R. (1977). Post-embryonic cell lineages of the nematode, Caenorhabditis elegans. *Developmental Biology*, *56*(1), 110–156. https://doi.org/10.1016/0012-1606(77)90158-0

Sulston, J. E., Schierenberg, E., White, J. G., & Thomson, J. N. (1983). The embryonic cell lineage of the nematode Caenorhabditis elegans. *Developmental Biology*, *100*(1), 64–119. https://doi.org/10.1016/0012-1606(83)90201-4

Tillberg, P. W., Chen, F., Piatkevich, K. D., Zhao, Y., Yu, C.-C. (Jay), English, B. P., Gao, L., Martorell, A., Suk, H.-J., Yoshida, F., DeGennaro, E. M., Roossien, D. H., Gong, G., Seneviratne, U., Tannenbaum, S. R., Desimone, R., Cai, D., & Boyden, E. S. (2016). Protein-retention expansion microscopy of cells and tissues labeled using standard fluorescent proteins and antibodies. *Nature Biotechnology*, *34*(9), 987–992. https://doi.org/10.1038/nbt.3625

Triller, A., Seitanidou, T., Franksson, O., & Korn, H. (1990). Size and shape of glycine receptor clusters in a central neuron exhibit a somato-dendritic gradient. *The New Biologist*, *2*(7), 637–641.

Truckenbrodt, S., Maidorn, M., Crzan, D., Wildhagen, H., Kabatas, S., & Rizzoli, S. O. (2018). X10 expansion microscopy enables 25-nm resolution on conventional microscopes. *EMBO Reports*, *19*(9), e45836. https://doi.org/10.15252/embr.201845836

Vangindertael, J., Beets, I., Rocha, S., Dedecker, P., Schoofs, L., Vanhoorelbeeke, K., Hofkens, J., & Mizuno, H. (2015). Super-resolution mapping of glutamate receptors in C. elegans by confocal correlated PALM. *Scientific Reports*, *5*, 13532. https://doi.org/10.1038/srep13532

Varady, A., & Distel, M. (2020). Non-neuromodulatory Optogenetic Tools in Zebrafish. *Frontiers in Cell and Developmental Biology*, 8, 418. https://doi.org/10.3389/fcell.2020.00418

Vicidomini, G., Bianchini, P., & Diaspro, A. (2018). STED super-resolved microscopy. *Nature Methods*, 15(3), 173–182. https://doi.org/10.1038/nmeth.4593

Wang, S., Shin, T. W., Yoder, H. B., McMillan, R. B., Su, H., Liu, Y., Zhang, C., Leung, K. S., Yin, P., Kiessling, L. L., & Boyden, E. S. (2024). Single-shot 20-fold expansion microscopy. Nature Methods, 1–7. https://doi.org/10.1038/s41592-024-02454-9

Wang, Y., Eddison, M., Fleishman, G., Weigert, M., Xu, S., Wang, T., Rokicki, K., Goina, C., Henry, F. E., Lemire, A. L., Schmidt, U., Yang, H., Svoboda, K., Myers, E. W., Saalfeld, S., Korff, W., Sternson, S. M., & Tillberg, P. W. (2021). EASI-FISH for thick tissue defines lateral hypothalamus spatio-molecular organization. *Cell*, 184(26), 6361-6377.e24. https://doi.org/10.1016/j.cell.2021.11.024

Werner, C., Sauer, M., & Geis, C. (2021). Super-resolving Microscopy in Neuroscience. *Chemical Reviews*, 121(19), 11971–12015. https://doi.org/10.1021/acs.chemrev.0c01174

White, J. G., Southgate, E., Thomson, J. N., & Brenner, S. (1986). The structure of the nervous system of the nematode Caenorhabditis elegans. *Philosophical Transactions of the Royal Society of London. Series B, Biological Sciences*, 314(1165), 1–340. https://doi.org/10.1098/rstb.1986.0056

Winding, M., Pedigo, B. D., Barnes, C. L., Patsolic, H. G., Park, Y., Kazimiers, T., Fushiki, A., Andrade, I. V., Khandelwal, A., Valdes-Aleman, J., Li, F., Randel, N., Barsotti, E., Correia, A., Fetter, R. D., Hartenstein, V., Priebe, C. E., Vogelstein, J. T., Cardona, A., & Zlatic, M. (2023). The connectome of an insect brain. *Science*, 379(6636), eadd9330. https://doi.org/10.1126/science.add9330

Witvliet, D., Mulcahy, B., Mitchell, J. K., Meirovitch, Y., Berger, D. R., Wu, Y., Liu, Y., Koh, W. X., Parvathala, R., Holmyard, D., Schalek, R. L., Shavit, N., Chisholm, A. D., Lichtman, J. W., Samuel, A. D. T., & Zhen, M. (2021). Connectomes across development reveal principles of brain maturation. *Nature*, 596(7871), 257–261. https://doi.org/10.1038/s41586-021-03778-8

Yao, C., Vanderpool, K. G., Delfiner, M., Eddy, V., Lucaci, A. G., Soto-Riveros, C., Yasumura, T., Rash, J. E., & Pereda, A. E. (2014). Electrical synaptic transmission in developing zebrafish: Properties and molecular composition of gap junctions at a central auditory synapse. *Journal of Neurophysiology*, 112(9), 2102–2113. https://doi.org/10.1152/jn.00397.2014

Yemini, E., Lin, A., Nejatbakhsh, A., Varol, E., Sun, R., Mena, G. E., Samuel, A. D. T., Paninski, L., Venkatachalam, V., & Hobert, O. (2021). NeuroPAL: A Multicolor Atlas for Whole-Brain Neuronal Identification in C. elegans. *Cell*, 184(1), 272-288.e11. https://doi.org/10.1016/j.cell.2020.12.012

Yoshimura, J., Ichikawa, K., Shoura, M. J., Artiles, K. L., Gabdank, I., Wahba, L., Smith, C. L., Edgley, M. L., Rougvie, A. E., Fire, A. Z., Morishita, S., & Schwarz, E. M. (2019). Recompleting the Caenorhabditis elegans genome. *Genome Research*, 29(6), 1009–1022. https://doi.org/10.1101/gr.244830.118

Yu, C.-C. (Jay), Barry, N. C., Wassie, A. T., Sinha, A., Bhattacharya, A., Asano, S., Zhang, C., Chen, F., Hobert, O., Goodman, M. B., Haspel, G., & Boyden, E. S. (2020). Expansion microscopy of C. elegans. *eLife*, 9, e46249. https://doi.org/10.7554/eLife.46249

Yu, C.-C., Orozco Cosio, D. M., & Boyden, E. S. (2022). ExCel: Super-Resolution Imaging of C. elegans with Expansion Microscopy. In G. Haspel & A. C. Hart (Eds.), *C. elegans* (Vol. 2468, pp. 141–203). Springer US. https://doi.org/10.1007/978-1-0716-2181-3_9

Zottoli, S. J., Hordes, A. R., & Faber, D. S. (1987). Localization of optic tectal input to the ventral dendrite of the goldfish Mauthner cell. *Brain Research*, *401*(1), 113–121. https://doi.org/10.1016/0006-8993(87)91170-X